# Nanoscale polarization manipulation and imaging in ferroelectric Langmuir-Blodgett polymer films

Brian J. Rodriguez[1], Stephen Jesse[1], Sergei V. Kalinin[1,*], Jihee Kim[2], Stephen Ducharme[2,**], & V. M. Fridkin[2,3]

[1]*Materials Science and Technology Division, Oak Ridge National Laboratory, Oak Ridge, TN 37831, USA*

[2]*Department of Physics and Astronomy, Nebraska Center for Materials and Nanostructures, University of Nebraska, Lincoln, NE 68588-0111, USA*

[3]*Institute of Crystallography, Russian Academy of Sciences, Moscow 117333, Russia*

*Email: sergei2@ornl.gov

**Email: sducharme@unl.edu

----

**The behavior of ferroelectricity at the nanoscale is the focus of increasing research activity because of intense interest in the fundamental nature of spontaneous order in condensed-matter systems and because of the many practical applications of ferroelectric thin films to, for example, electromechanical transducers,[1] infrared imaging sensors,[2] and nonvolatile random-access memories.[3] Ferroelectricity, in analogy with its namesake ferromagnetism, is the property of some crystalline systems to maintain a permanent, but reversible, electrical polarization in the absence of an external electric field. The imaging and dynamics of the piezoelectric response at the nanoscale is perhaps the most direct means of probing polarization, as has been demonstrated in a number of thin films[4] and nanostructures[5]. Here we report the use of piezoresponse force microscopy (PFM) and switching spectroscopy PFM (SSPFM) to image the ferroelectric properties, domain structure, and polarization switching of ultrathin ferroelectric Langmuir-Blodgett films of poly(vinylidene fluoride-trifluoroethylene) (P(VDF-TrFE)) copolymers. PFM imaging of P(VDF-TrFE) thin films reveals ferroelectric domain sizes on the order of 25-50 nm and an imaging resolution below 5 nm. The feature sizes in topography and PFM images are comparable and the boundaries of uniformly polarized regions coincide with topographic features. Arbitrary polarization patterns could be repeatedly written and erased, with writing resolution limited by the grain size. Hysteresis loops from individual domains show clear coercive voltage, but are not well saturated at ±10 V amplitude.**

The ferroelectric polymer, polyvinylidene fluoride (PVDF), is being developed for use in nonvolatile data storage,[6] and flexible ferroelectric electronic components due to its outstanding electromechanical, dielectric, and mechanical properties.[7,8,9] High-quality thin films of PVDF and its copolymers can be fabricated using a Langmuir-Blodgett (LB) technique. These 2-dimensional ferroelectric films have allowed the intrinsic coercive field to be measured.[10] In order to realize the full potential of PVDF for applications such as ultrahigh-density non-volatile memories, significant progress must be made in nanoscale characterization of the structure and ferroelectric properties of PVDF LB films, including local hysteresis and switching dynamics.

Polyvinylidene fluoride has been the focus of numerous scanning probe microscopy investigations. Güthner and Dransfeld pioneered the use of scanning near field acoustic microscopy,[11] scanning tunneling microscopy,[12] and atomic force microscopy[13] (AFM) for electromechanical imaging and polarization switching in spin-coated copolymer films of polyvinylidene fluoride and trifluoroethylene (P(VDF-TrFE)). Particularly useful for studies of ferroelectric materials is piezoresponse force microscopy (PFM), a kind of electromechanical imaging achieved by applying an ac bias to a conducting tip in a contact-mode AFM and detecting surface displacement at the frequency of ac modulation. PFM is routinely conducted on ferroelectric crystals, films, and nanostructures,[14,15,16] and several recent PFM studies of ferroelectric polymers[17,18] and oligomers[19,20] have been reported. Spin-coated thin films of P(VDF-TrFE) have been poled and imaged by PFM and ferroelectric domains as small as 30 nm were created.[17,18] In addition, an asymmetric piezoelectric behavior was observed and attributed to a frozen layer at the substrate interface.[17] Atomic-resolution imaging of P(VDF-TrFE) monolayers by scanning tunneling microscopy (STM) has produced images of discontinuities in molecular conformation, which have been interpreted as domain walls perpendicular to the chain,[21] but these studies did not determine local polarization. PFM has also been used to image polarization structure and local switching in P(VDF-TrFE) LB films[22] and nanotubes (Y. Luo *et al.*, in preparation). Electrostatic force microscopy has been used to investigate the ferroelectric properties of LB P(VDF-TrFE) thin films, however no domain structure was observed within crystals.[23] In this letter, we report on nanoscale polarization imaging and polarization switching in LB films of P(VDF-TrFE), accomplished by PFM, showing imaging resolution of 5 nm and switchable polarization in regions averaging 25-50 nm in size.

The thin ferroelectric P(VDF-TrFE) copolymer films were fabricated on highly ordered pyrolytic graphite (HOPG) substrates by an LB technique.[24] Deionized water with a resistivity of 18 MΩ was used as a pure water subphase and placed in a clean teflon trough with a total area of 1520 cm$^2$. Then, a solution of vinylidene fluoride (70%) with trifluoroethylene (30%) copolymer, P(VDF-TrFE 70:30), in dimethyl sulfoxide (0.01% weight concentration) was dispersed on the top of the water subphase with the aid of an electric pipette and microslides. The dispersed copolymer solution became a thin film on the water subphase, and the film was compressed at a rate of 20~60 cm$^2$/min by two barriers from the outside towards the center of the trough to reach a target surface pressure of 5 mN/m. Once the target pressure was reached, the film on the water surface was transferred to the HOPG substrate by horizontal dipping. The water trough was kept at 25 ºC during entire deposition process. Samples were then annealed for 1 hour at 130 ºC to optimize their crystal structure and stabilize the ferroelectric properties.[25] Prior studies showed that this technique produces films with and average thickness of 1.78±0.07 nm per nominal monolayer (ML) transferred.[26]

Piezoresponse force microscopy employs the converse piezoelectric effect to measure small surface displacements resulting from the application of an external ac field at the contact between a conducting tip and the sample surface. The tip follows the expansion and contraction of the surface allowing the voltage-dependent piezoelectric response to be mapped simultaneously with topography using a lock-in technique. A schematic diagram of the PFM measurement is shown as an inset in Fig. 1(e), and a more comprehensive description is provided elsewhere.[4] Piezoresponse force microscopy and spectroscopy studies were implemented on a commercial AFM system (Veeco MultiMode NS-IIIA) equipped with an external signal generation and data acquisition system.[27] Measurements were performed using Au-Cr coated Si tips (Micromasch, spring constant $k \sim 0.08$ N/m).

To determine the structural ordering of ferroelectricity in 10 ML films, PFM was used. Topography and PFM amplitude ($d_{1\omega}$) and phase ($\varphi$) images are shown in Figure 1 (a, b, and c, respectively). The AFM topography image (Fig. 1a) shows that the LB film is highly planar, with a root-mean-square roughness of 1.4±0.2 nm. The PFM amplitude image (Fig. 1b) shows irregularly-shaped regions of uniform piezoelectric response, outlined by narrow unpolarized regions that appear to be intergrain boundaries, although this is less evident in the topography image (Fig. 1a) due to the loading force associated with contact mode imaging. In the PFM phase image (Fig 1c), a bright region corresponds to a single domain that has a polarization oriented toward the substrate, while a dark region corresponds to a region with polarization oriented away from the substrate, a 'down' polarization and an 'up' polarization, respectively. Boundaries can be seen in the PFM amplitude image (Fig 1b) between two opposite polarizations (one such domain wall is indicated with an arrow in Fig. 1b). This minimum in the amplitude signal corresponds to an apparently unpolarized boundary wall separating two grains. There is a distribution of topographic feature sizes between ~25-50 nm in diameter. For a polarization oriented perpendicular to the substrate, a field applied parallel to the dipole will result in a piezoresponse perpendicular to the substrate.

To illustrate imaging resolution we generated a composite image (Fig. 1d) of the mixed piezoresponse, $PR = d_{1\omega} \cos\varphi$, from the amplitude and phase images (Figs. 1b,c). The composite image clearly shows domains of opposing polarization as bright 'up' polarization and dark 'down' polarization. A line profile across one of these grain boundaries is shown in Fig. 1(e). The width of the transition from positive to negative piezoresponse determines the imaging resolution, approximately 5 nm in this case, an order of magnitude better than is typically observed for inorganic ferroelectrics.[17, 18] The high resolution is achieved because the LB films are highly planar, without complex lamellar morphologies that dominate solvent-formed films.

To investigate the polarization switching of LB films of P(VDF-TrFE), a dc bias was applied while scanning in order to switch the polarization from its as-annealed state. First, a square area was poled in one direction by the application of a dc bias while scanning 256 lines at 1 line per second. Then, a smaller square area was poled in the opposite direction by the application of a dc bias of opposite polarity, and the entire region was imaged with an ac bias. In Fig. 2(a, b, c) topography, PFM amplitude, and PFM phase images, respectively, are shown following poling with +/- 10 V dc bias. In Fig. 2(d, e, f), a larger area has been poled and imaged with the same conditions. As can be seen in the PFM images in Fig. 2(b, c, e, f), the regions of uniform polarization are rough and irregular, and correspond to well to the topographic structure, suggesting that switching occurs one grain at a time, rather than on a multiple grain basis. The less uniform pattern in Fig. 2 (b,c) is indicative of a degraded tip. Domains written using a square wave bias (+/- 10 V) pattern along the scanning direction produced the series of parallel lines shown in Fig. 2 (g, h, i). Interestingly, even though dipole rotation and conformation changes are inseparable given the covalent bonding of the chains, it is still possible to switch a single grain without rotating the polarization of nearby grains when the field is applied locally using a conductive AFM tip. The polarization direction of the grains of annealed P(VDF-TrFE) LB films can be determined and switched with a writing resolution corresponding to the grain size and an imaging resolution of 5 nm.

To investigate the local switching properties in more detail, switching spectroscopy PFM (SSPFM) was employed, in which a local hysteresis loop was measured at every point within a specified grid.[27] The switching parameters (nucleation bias, imprint, remanent piezoresponse, etc.) can then be extracted from the loops and plotted as 2D maps. In this manner, the dynamics of the local switching characteristics can be probed and the switching properties can be visualized to show how these characteristics vary across a sample surface.[28] In Fig. 3 an SSPFM image of the negative remanent piezoresponse within a 2 x 2 $\mu m^2$ area and representative hysteresis loops demonstrating a variation in the vertical shift are shown. The hysteresis loops measured on P(VDF-TrFE) do not saturate with the range ±10 V and attempts to bring the loops to saturation by increasing the dc bias typically ended with dielectric breakdown at biases of 10 V and higher depending on the location, precluding a systematic study of the switching properties.

In summary, PFM studies of P(VDF-TrFE) thin films demonstrate nanoscale imaging and control of polarization in ultrathin ferroelectric polymer films. The imaging resolution was better than 5 nm and images showed single-domain regions corresponding to feature sizes averaging 25-50nm; the observed polarization patterns follow the irregular grain topography and the grains switch independently of each other. Arbitrary polarization patterns could be repeatedly written and erased with a resolution limited by the grain size, showing that these films are suitable for use in high-density nonvolatile memories and ferroelectric devices, ultimately down to the molecular level.

**Acknowledgements.** Work at the Oak Ridge National Laboratory was supported by ORNL SEED funding under contract DE-AC05-00OR22725 (B.J.R., S.J., S.V.K.). Work at the University of Nebraska (J.K., S.D., V.M.F.) was supported by the National Science Foundation and the Nebraska Research Initiative. The authors are grateful to V. Meunier for insightful discussions.

**Figure Captions**

**Fig. 1.** PFM data from a 10-ML Langmuir-Blodgett film of P(VDF-TrFE 70:30) copolymer: (a) topography; (b) PFM amplitude; (c) PFM phase; (d) composite (amplitude*cos(phase)) PFM; (e) line scan of the composite image along the green bar. The scale for images a-c is the same.

**Fig. 2.** PFM images of patterns that were written with a tip bias of ±10 V in a 10-ML Langmuir-Blodgett film of P(VDF-TrFE 70:30) copolymer: (a,d,g) topography; (b,e,h) PFM amplitude; (c,f,i) PFM phase. Patterns were written with +/- 10V. The quality of the switched pattern depends on the voltage, the film, and also the tip-state. The domain walls are irregular and correspond to topographic features.

**Fig. 3.** Polarization hysteresis loops obtained from a 10-ML Langmuir-Blodgett film of P(VDF-TrFE 70:30) copolymer: (a) SSPFM image of the negative remanent piezoresponse for a 2 x 2 $\mu m^2$ region (32 nm step size), and (b) representative hysteresis loops from regions indicated in (a).

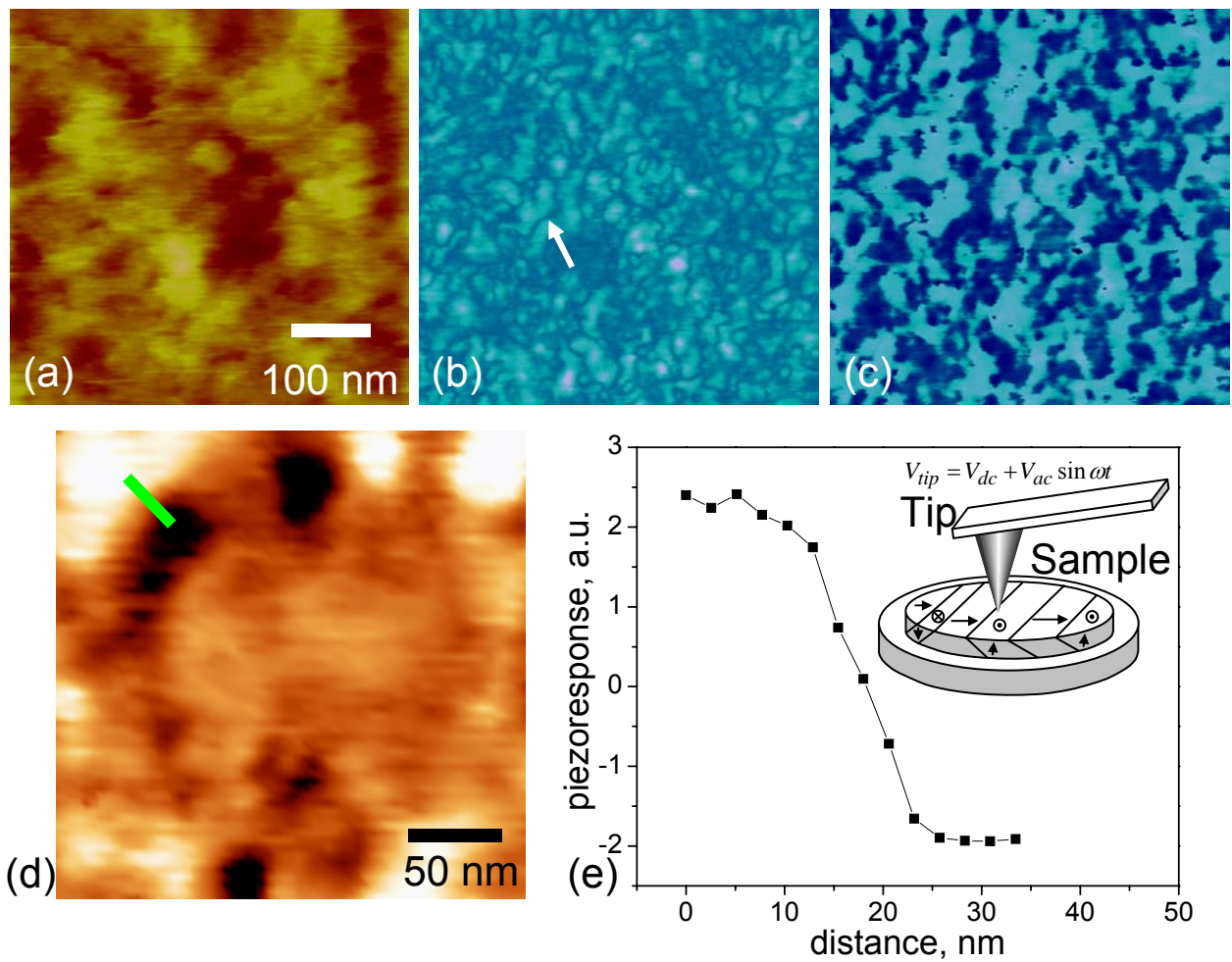

**Figure 1.** B. J. Rodriguez, S. Jesse, S. V. Kalinin, J. Kim, S. Ducharme, V. Fridkin

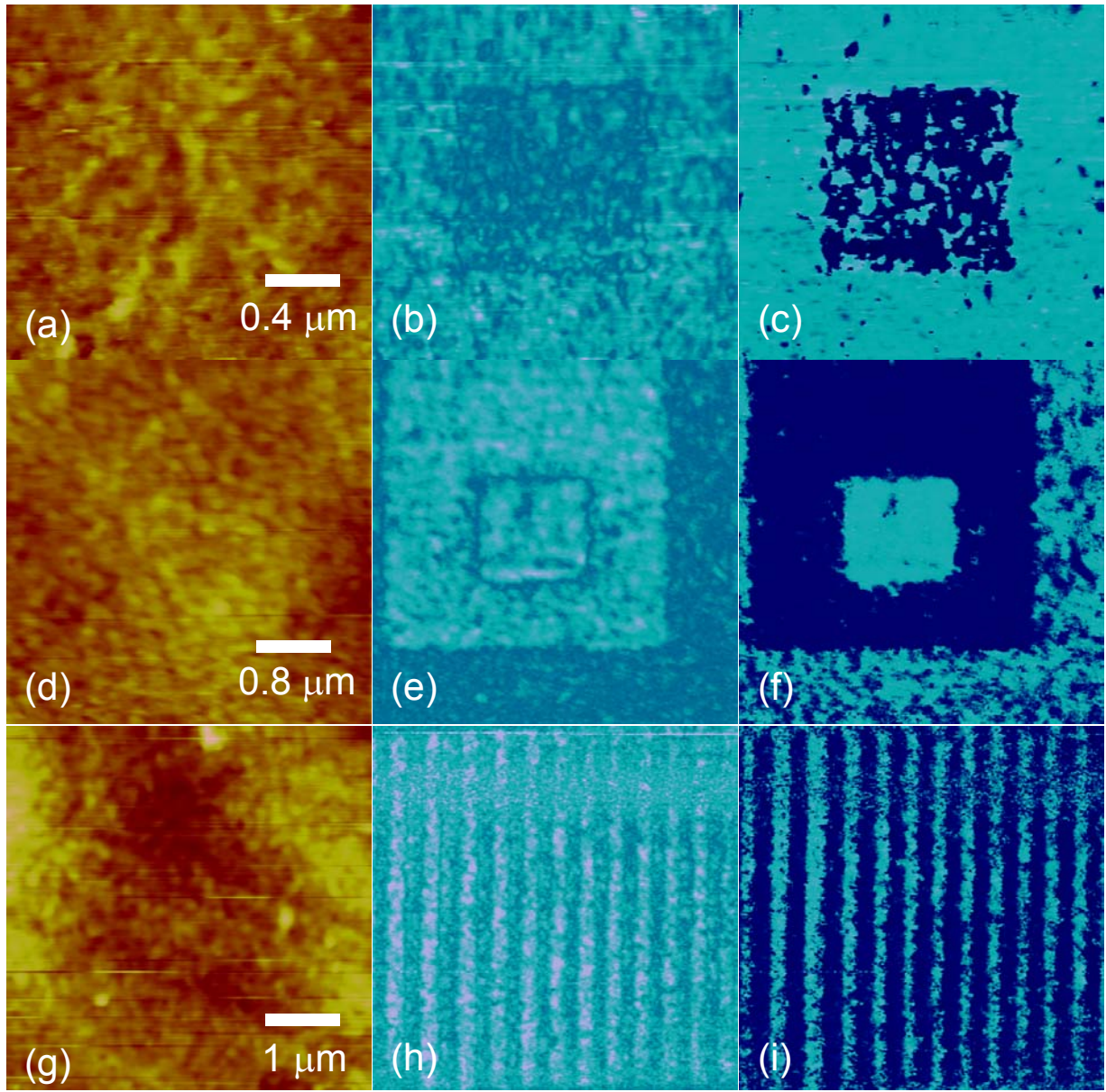

**Figure 2.** B. J. Rodriguez, S. Jesse, S. V. Kalinin, J. Kim, S. Ducharme, V. Fridkin

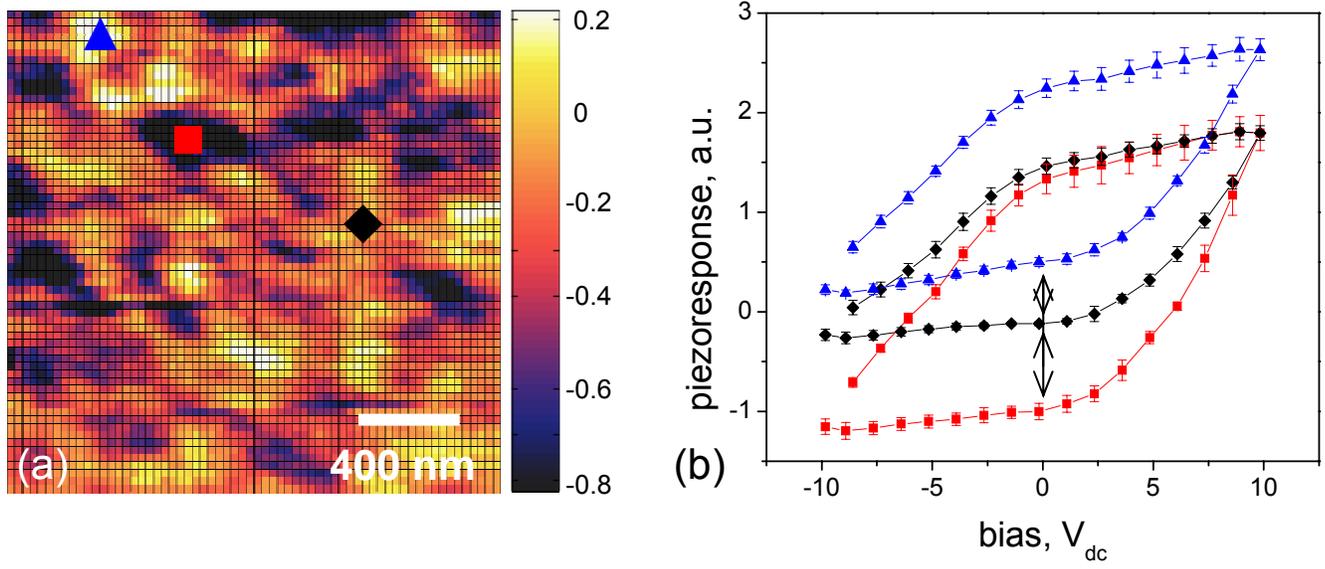

**Figure 3.** B. J. Rodriguez, S. Jesse, S. V. Kalinin, J. Kim, S. Ducharme, V. Fridkin